\documentclass[aps,prd,onecolumn,groupedaddress,showpacs,nofootinbib,amssymb]{revtex4}
\usepackage{graphicx,bm,color}
\usepackage{amsmath}
\usepackage{amssymb}
\usepackage{amsfonts}
\usepackage{ulem}

\newcommand{\be}{\begin{equation}}
\newcommand{\ee}{\end{equation}}
\newcommand{\bea}{\begin{eqnarray}}
\newcommand{\eea}{\end{eqnarray}}
\newcommand{\beaa}{\begin{eqnarray*}}
\newcommand{\eeaa}{\end{eqnarray*}}

\newcommand{\nn}{\nonumber \\}
\newcommand{\e}{\mathrm{e}}

\allowdisplaybreaks[4]

\begin{document}

\title{Palatini-Born-Infeld Gravity, Bouncing Universe, and Black Hole Formation}

\author{Meguru Komada$^1$, Shin'ichi Nojiri$^{1,2}$, and Taishi Katsuragawa$^1$}

\affiliation{
1. Department of Physics, Nagoya University, Nagoya
464-8602, Japan \\
2. Kobayashi-Maskawa Institute for the Origin of Particles and
the Universe, Nagoya University, Nagoya 464-8602, Japan 
}

\begin{abstract}

We consider the Palatini formalism of the Born-Infeld gravity. In the Palatini formalism, 
the propagating mode is only graviton, whose situation is different from that in the metric 
formalism. We discuss about the FRW cosmology by using an effective potential. Especially we consider the 
condition that the bouncing could occur. 
We also give some speculations about the black hole formation 

\end{abstract}


\maketitle

\section{Introduction}

In recent years, many kinds of modified gravity theories have been proposed and 
investigated in several motivations.
First motivation could be the quantum gravity.
We have not obtained any consistent quantum theory of the gravity. 
Although Einstein's general relativity is a successful theory of the classical theory of gravity, 
we cannot obtain the quantum theory based on the general relativity due to the non-renormalizability. 
Motivated with the quantum gravity, many kinds of modification of the Einstein gravity have been 
proposed and investigated. 
On the other hand, motivated with the accelerating expansion of the present universe, 
we are considering many kinds of gravity theories beyond the Einstein gravity 
(for review, see \cite{review}). 
As a model of such modified gravities, in this paper, we consider about the Born-Infeld 
gravity \cite{Deser:1998rj} in the Palatini  formalism \cite{Vollick:2003qp}. 
The Born-Infeld type theory was first proposed as a non-linear model of electro-magnetics 
\cite{Born:1934gh}. 
In the Maxwell electro-magnetics, the action of the electromagnetic field has no 
dimensional parameter but in the Born-Infeld model, a scale can be introduced. 
Because the action includes the 
square root, there appears the upper limit in the strength given by the scale, which may have suggested 
that there might not appear the divergence in the quantum field theory different from the standard  
quantum electro-dynamics. 
The Born-Infeld type deformation was also considered in gravity theory \cite{Deser:1998rj} and 
it has been expected that the quantum theory of the Born-Infeld gravity might be finite and 
there might not appear any divergence because there could be an upper limit in the magnitude of the 
curvature. 
By including the idea by Eddington \cite{Eddington}, the Born-Infeld gravity has been formulated in 
the Palatini formalism, where the connections are variables independent from the metric tensor. 
The corresponding action is given by
\be
\label{PBIG1}
S = \frac{1}{\kappa^2 b} \int d^4 x \left\{ \sqrt{ \left| \det \left( g_{\mu\nu} + b R_{\mu\nu} 
\right) \right| } - \sqrt{ \left| \det \left( g_{\mu\nu} \right) \right| } \right\}
+ S_\mathrm{matter} \, .
\ee
Here $\kappa$ is the gravitational constant corresponding to the Einstein gravity and we introduce a new 
parameter $b$, which has the dimension of the square of the length. 
In Eq.~(\ref{PBIG1}), $S_\mathrm{matter}$ is an action for the matters and 
$R_{\mu\nu}$ is the Ricci tensor assumed to be symmetric and defined by 
$R_{\mu\nu} = -\Gamma^\rho_{\mu\rho,\nu} + \Gamma^\rho_{\mu\nu,\rho}
- \Gamma^\eta_{\mu\rho}\Gamma^\rho_{\nu\eta} + \Gamma^\eta_{\mu\nu}\Gamma^\rho_{\rho\eta}$ 
in terms of the connection $\Gamma^\rho_{\mu\nu}$. 
We regard the connection $\Gamma^\rho_{\mu\nu}$ is a variable independent from the metric 
$g_{\mu\nu}$. The theory described by the above action (\ref{PBIG1}) is often called the Eddington inspired 
Born-Infeld gravity.
By the variations of the action with respect to the metric $g_{\mu\nu}$ and the connection 
$\Gamma^\lambda_{\mu\nu}$, we obtain the following equations, respectively, 
\begin{align}
\label{eq:metric}
0 = & \sqrt{-P} \left(P^{-1} \right)^{\mu \nu } - \sqrt{-g} g^{\mu \nu} 
 - b\kappa^2 \sqrt{-g} T^{\mu\nu}
\, , \\
0 = &\nabla_{\lambda} \left( \sqrt{-P} \left(P^{-1}\right)^{\mu \nu }\right) = 0\, .
\label{eq:connection}
\end{align}
Here $T^{\mu\nu}$ is the energy-momentum tensor of the matters and $P_{\mu\nu}$ is defined by
\be
\label{PBIG2}
P_{\mu \nu } \equiv g_{\mu \nu } + bR_{\mu \nu } \, ,
\ee
and $P^{-1}$ is the inverse of the matrix $P_{\mu\nu}$, that is, 
$\left(P^{-1}\right)^{\mu\rho} P_{\rho\nu} = \delta^\mu_{\ \nu}$. 
Then $P^{-1}$ can be expressed by the infinite power series of $R_{\mu\nu}$, 
$\left(P^{-1}\right)^{\mu\nu} = g^{\mu\nu} - b g^{\mu\rho} R_{\rho\sigma} g^{\sigma\nu} 
+  b^2 g^{\mu\rho} R_{\rho\sigma} g^{\sigma\tau} R_{\tau\eta} g^{\eta\nu} - \cdots$.  
On the other hand $P^{\mu\nu}$ is defined by $P^{\mu\nu} \equiv g^{\mu\rho} P_{\rho\sigma} g^{\sigma\nu} = g^{\mu\nu} + b R^{\mu\nu}$.

In the metric formalism of the Born-Infeld gravity, theory includes a ghost 
in general and we need to tune the action by adding the higher derivative terms so that the ghost does 
not appear \cite{Deser:1998rj}. 
In the Palatini formalism, however, there does not appear ghost \cite{Vollick:2003qp} and 
the only propagating mode is massless graviton. 
Then in the leading order, the standard Newton law can be reproduced although there also appear some 
corrections \cite{Banados:2010ix}. 
We should note that the Born-Infeld gravity in the Palatini formalism is equivalent to the 
Einstein gravity in the vacuum although they are different from each other when the matter exists.

The cosmology by the Born-Infeld gravity in the Palatini formalism has been 
considered in \cite{Banados:2010ix,Avelino:2012ue} by including matters. 
The development of the universe in the Born-Infeld gravity in the Palatini formalism
shows the behavior which is different from that in the Einstein gravity and there often appears the 
bouncing universe, where the shrinking universe turns to expand. 
We claim, however, that the treatments in the previous papers were not complete and 
we re-examine the cosmology and we show that there anyway appears the bouncing universe as a solution.

Even in the Palatini-Born-Infeld gravity, the Schwarzschild space-time and the Kerr black hole 
space-time are exact solutions. 
Because these solutions are vacuum solution, they are equivalent with the solutions in the 
Einstein gravity and therefore, for example, the expressions of the black hole entropies are identical 
with those in the Einstein gravity (in case of the metric formalism, see \cite{Chemissany:2008fy}).

In this paper, we investigate the FRW cosmology.
We should note that in the previous works, there were too strong 
constraints on the variables but we consider more general treatment in this paper. 
We consider the cosmology by including dust as a matter and show that the 
bouncing universe can be realized, whose behavior is, in some sense, similar to that in the 
loop quantum gravity \cite{Bojowald:1999tr,Bojowald:1999ts,Ashtekar:2006es}. 
We also give some speculations about the formation of the black hole by considering the collapse of the 
sphere of the dust. 
Because the pressure of the dust vanishes, we can regard the inside of the sphere as 
the FRW universe. Then by using the results in the FRW universe, we show that the small 
black hole could not be formed by the bouncing although the large black holes might be created. 
We should note, however, we do not use the junction conditions in the Born-Infeld gravity in the Palatini 
formalism but we use those in the Einstein gravity. 
This is because the junction conditions in the Born-Infeld gravity in the Palatini formalism are very 
complicated and, to be honest, we have not found the full expressions of the junction conditions. 
We will discuss this point in more detail later.
Therefore the analyses in this paper could not be justified in some cases but by using the obtained 
results, we can give some realistic speculations.

\section{FRW universe with dust}

We consider the FRW space-time with flat spacial part, 
\be
\label{MK1}
ds^2 = - dt^2 + a (t)^2 \sum_{i=1,2,3} \left( dx^i \right)^2 \, ,
\ee
and assume that the non-vanishing components of the connection are given by
\be
\label{MK2}
\Gamma^t_{tt} = A (t)\, ,\quad \Gamma^t_{ij} = a(t)^2 B(t) \delta{ij}\, , \quad 
\Gamma^i_{jt} = \Gamma^i_{tj} = C(t) \delta^i_{\ j}\, .
\ee
In the Einstein gravity, 
the metric (\ref{MK2}) implies 
$A=0$, $B=C=H \equiv \dot a / a$. 

In the previous works \cite{Banados:2010ix,Avelino:2012ue}, the FRW metric was assumed for $P_{\mu \nu }$ in Eq.~(\ref{PBIG2}),  
\be
\label{PP1}
ds_P^2 = \sum_{\mu,\nu=0}^4 P_{\mu\nu} dx^\mu dx^\nu 
= - dt^2 + \tilde a (t)^2 \sum_{i=1,2,3} \left( dx^i \right)^2 \, ,
\ee
in \cite{Banados:2010ix} and 
\be
\label{PP1b}
ds_P^2 = \sum_{\mu,\nu=0}^4 P_{\mu\nu} dx^\mu dx^\nu 
= - \tilde b (t)^2 dt^2 + \tilde a (t)^2 \sum_{i=1,2,3} \left( dx^i \right)^2 \, ,
\ee
in \cite{Avelino:2012ue} and the connection $\Gamma^\lambda_{\mu\nu}$ was given by $P_{\mu\nu}$:
\be
\label{PP2}
\Gamma^\lambda_{\mu\nu}
= \frac{1}{2} \left(P^{-1}\right)^{\lambda\rho} 
\left( \partial_\mu P_{\rho\nu}
+ \partial_\nu P_{\mu\rho} - \partial_\rho P_{\mu\nu} \right)\, ,
\ee
which, however, reduces the degrees of freedom in  $\Gamma^\lambda_{\mu\nu}$. 
We should note that $P_{\mu\nu}$ is not the fundamental variable in the Palatini formalism but 
$\Gamma^\lambda_{\mu\nu}$ is the fundamental one. 
As we find in Eq.~(\ref{MK2}), even if we assume homogeneous and isotropic universe, there appear three 
undetermined variables $A$, $B$, and $C$ but in Eq.~(\ref{PP1}), only one undetermined variable $\tilde a$ 
appears and in Eq.~(\ref{PP2}), two variables $\tilde b$ and $\tilde a$. 
Therefore the assumption (\ref{PP1}) or (\ref{PP2}) may conflict with the equations of the Born-Infeld 
gravity in the Palatini formalism. 
In fact, if we assume Eq.~(\ref{PP1}), we find $A=0$ and $B=C$, which conflict with the analysis given later in this paper.
Even if we assume Eq.~(\ref{PP1b}), there occur conflictions.\footnote{
Even in case of Eq.~(\ref{PP1b}), we find $A=\dot{\tilde b} / \tilde b$, $B=\dot{\tilde a}/a b^2$, $C=\dot{\tilde a}/{a}$, 
which satisfy the relation $2 A = \dot C/C - \dot B/B$. If we use this relation for Eqs.~(\ref{MK9}), (\ref{MKSN5}), 
and (\ref{MKSN7}) with Eq.~(\ref{MKSN2}), we obtain
\[
b\kappa^2 \rho = -8\, ,
\]
which is inconsistent.
Furthermore we should note that the difference between Eqs.~(\ref{PP1}) and (\ref{PP2}) is nothing but the parametrization of the time coordinate. 
} 
Although there might be cases that the assumption (\ref{PP1}) or (\ref{PP1b}) could be justified, we do 
not know any a priori reason. 
Therefore we like to re-investigate the cosmology under the assumptions (\ref{MK1}) and (\ref{MK2}). 

We now assume that the matter is given by the dust whose pressure $p$ vanishes and the energy 
density is denoted by $\rho$. 
One of the reasons why we consider the dust as the matter is for the simplicity although it is not so difficult if we consider general perfect fluid as the matter. Another reason is because we like to compare the black hole formation in this model with the Oppenheimer-Snyder collapse in the general relativity \cite{Oppenheimer:1939}, where the inside of the collapsing star is described by the FRW universe filled with dust. 
Because the dust evolves in the space-time whose metric is given by $g_{\mu\nu}$, 
we obtain the conservation law identical with that in the Einstein gravity, 
$\nabla^{(g)\mu} T_{\mu\nu} = 0$. 
Here $\nabla^{(g)\mu}$ is the covariant derivative where the connection is given in terms of 
$g_{\mu\nu}$ as in the standard Einstein gravity but not by $P_{\mu\nu}$. 
Then in the FRW universe (\ref{MK1}), we obtain the standard conservation law 
\be
\label{MKSN2}
\dot \rho + 3H p = 0 \, .
\ee
and we find  
\be
\label{MKSN1}
\rho = \rho_0 a^{-3}\, .
\ee
Then we obtain the following equations (the derivations of the following equations are given in 
Appendix \ref{derivation}):
\begin{align}
\label{MK9}
A =&C-H= - \frac{3}{4} b \kappa^2 H \rho \left( 1 + b \kappa^2 \rho \right)^{-1}\, , \\
\label{MKSN5}
B =& H + \frac{b\kappa^2}{4} H \rho\, , \\
\label{MKSN7}
C =& H - \frac{3}{4} b \kappa^2 H \rho \left( 1 + b \kappa^2 \rho \right)^{-1}\, , \\
\label{MKSN8}
b \kappa^2 \rho &= \left\{ 1 + b \left( \dot H + 3 H^2 
+ \frac{b\kappa^2}{2} \dot H \rho \right) \right\}^2 - 1 \, .
\end{align}
When $b<0$, Eq.~(\ref{MKSN8}) tells that there is an upper limit $\rho_\mathrm{u}$ for the energy 
density $\rho$: 
\be
\label{MK14}
\rho_\mathrm{u} = - \frac{1}{b \kappa^2}\, .
\ee
Because $H=\dot a/a$ and $\dot H = \ddot a / a - {\dot a}^2 / a^2$, by using Eq.~(\ref{MKSN1}), 
Eq.~(\ref{MKSN8}) can be rewritten as 
\be
\label{MKSN9}
b \kappa^2 \rho_0 a^{-3} = \left\{ 1 + b \left( \frac{ \ddot a}{a} 
+ 2 \left( \frac{\dot a}{a} \right)^2  + \frac{b\kappa^2}{4} \left( \frac{\ddot a}{a} 
 - \left( \frac{\dot a}{a} \right)^2 \right) \rho_0 a^{-3} \right) \right\}^2 - 1 \, ,
\ee
which is a single equation with respect to the scale factor $a$. 
If we use the e-foldings $N$ defined by $a=\e^N$, we obtain
\be
\label{MKSN10}
b \kappa^2 \rho_0 \e^{-3N} = \left\{ 1 + b \left( \ddot N + 3 {\dot N}^2 
+ \frac{b\kappa^2}{4} \ddot N \rho_0 \e^{-3N} \right) \right\}^2 - 1 \, ,
\ee
or
\be
\label{MKSN11}
\ddot N = - \frac{ 3 {\dot N}^2 }{  1 + \frac{b\kappa^2 \rho_0}{4} \e^{-3N}} 
  - \frac{1 - \sqrt{ 1 + b \kappa^2 \rho_0 \e^{-3N} }}
{b \left( 1 + \frac{b\kappa^2 \rho_0}{4} \e^{-3N}\right) } \, .
\ee
By an analogy with the Newton equation in the classical mechanics, 
the first term in the r.h.s. could be an analogue of the drag force in the fluid when the Reynolds number is large 
and the second term could be a force by a potential, which we denote by $F(N)$ as
\be
\label{MKSN12}
F(N) \equiv   - \frac{1 - \sqrt{ 1 + b \kappa^2 \rho_0 \e^{-3N} }}
{b \left( 1 + \frac{b\kappa^2 \rho_0}{4} \e^{-3N}\right) } \, .
\ee
We should note that the potential force $F(N)$ is positive, which does not depend on 
the sign of the parameter $b$ and therefore the force acts so that the e-foldings $N$ 
increases.   
Then even if the universe is shrinking, it may turn to expand. 
Because 
\be
\label{MKSN13}
F'(N) \equiv \frac{ 3 \kappa^2 \rho_0 \e^{-3N}\left( 1 - \frac{b \kappa^2 \rho_0}{2} \e^{-3N} 
+ \sqrt{ 1 + b \kappa^2 \rho_0 \e^{-3N} } \right)}{4 \left( 1 
+ \frac{b\kappa^2 \rho_0}{4} \e^{-3N}\right)^2 \sqrt{ 1 + b \kappa^2 \rho_0 \e^{-3N} }} \, ,
\ee
we find, 
\begin{itemize}
\item When $b>0$, there is a maximum $F(N)= 2/3b$ at $b \kappa^2 \rho_0 \e^{-3N}=8$. 
We also find that $F(N)\to 0$ when $N\to + \infty$ and $F(N)\to 0$ when $N\to - \infty$.
\item When $b<0$, we find $F'(N)<0$ and therefore there is a maximum $F(N) = - 4/3b$ at 
$b \kappa^2 \rho_0 \e^{-3N}= -1$ and $F(N)\to 0$ when $N\to + \infty$, again. 
\end{itemize}

Eq.~(\ref{MKSN11}) can be further rewritten in the following form:
\be
\label{MKSN16}
0 = \frac{d^2}{dt^2} \left( \frac{\e^{3N}}{3} + \frac{b\kappa^2 \rho_0}{4} N \right) 
+ \frac{\e^{3N} \left( 1 - \sqrt{ 1 + b \kappa^2 \rho_0 \e^{-3N} } \right) }{b}\, , 
\ee
which tells that there is a conserved quantity $E$, 
\begin{align}
\label{MKSN17}
E =& \frac{1}{2} \left\{ \frac{d}{dt} \left( \frac{\e^{3N}}{3} 
+ \frac{b\kappa^2 \rho_0}{4} N \right)\right\}^2 
+ \int^N dN \frac{ \e^{3N} \left( 1 - \sqrt{ 1 + b \kappa^2 \rho_0 \e^{-3N}} \right) 
\left( \e^{3N}  + \frac{b \kappa^2}{4} \rho_0 \right) }{b} \nn
=& \frac{1}{2} \left\{ \frac{d}{dt} \left( \frac{\e^{3N}}{3} 
+ \frac{b\kappa^2 \rho_0}{4} N \right)\right\}^2 + V(N) \, ,
\end{align}
which corresponds to the total energy in the classical mechanics. 
Here $V(N)$ is given by 
\begin{align}
\label{MKSN18}
V(N) =& \frac{\e^{6N}}{6b}\left( 1 + \frac{1}{2} b\kappa^2 \rho_0 \e^{-3N} \right)
\left( 1 - \sqrt{ 1 + b \kappa^2 \rho_0 \e^{-3N}} \right)
 - \frac{b\kappa^2 \rho_0}{12b} \e^{3N}  \sqrt{ 1 + b \kappa^2 \rho_0 \e^{-3N}} \, .
\end{align}
When $N$ is positive and large, $V(N)$ behaves as 
\be
\label{MKSN19}
V(N) \sim  - \frac{\kappa^2 \rho_0 \e^{3N}}{6} \, .
\ee
In case $b>0$, when $N$ is negative and large, we find 
\be
\label{MKSN20}
V(N) \sim - \frac{ \left( b \kappa^2 \rho_0 \right)^{\frac{3}{2}} \e^{\frac{3}{2}N}}{6b}\, .
\ee
On the other hand, in case $b<0$, there is a maximum in $V(N)$ when 
$1 + b \kappa^2 \rho_0 \e^{-3N}=0$: 
\be
\label{MKSN21}
V(N) = V_\mathrm{max} \equiv \frac{ \left( b\kappa^2 \rho_0 \right)^2 }{12b}<0\, .
\ee
We now assume that the universe may have started from $N\to + \infty$ and 
after that they have started to shrink. 
Then from the above results, by the analogy with the classical mechanics, we find the followings: 
\begin{itemize} 
\item In case $b>0$, if $E<0$, the shrinking of the universe will stop and turn to 
expand. On the other hand if $E>0$, the universe will continue to shrink and 
the scale factor $a$ vanishes in the infinite future. 
\item In case $b<0$, if $E<V_\mathrm{max}$, the shrinking of the universe will 
stop and turn to expand. On the other hand if $E>V_\mathrm{max}$, the universe 
will reach the singular point at $1 + b \kappa^2 \rho_0 \e^{-3N}=0$. 
\end{itemize}
In order to estimate $E$, we now solve Eq.~(\ref{MKSN11}) by assuming $N\gg 1$. 
Then Eq.~(\ref{MKSN11}) can be rewritten as 
\be
\label{MKSN22}
\ddot N + 3 {\dot N}^2 - \frac{\kappa^2 \rho_0 \e^{-3N}}{2}
= \frac{3}{4} b\kappa^2 \rho_0 \e^{-3N} {\dot N}^2 
 - \frac{ \left( b\kappa^2 \rho_0\right)^2 }{4b} \e^{-6N}  
+ \mathcal{O} \left( b^2 \right)\, .
\ee
Then in the limit $b\to 0$, we find 
\be
\label{MKSN23}
N = \frac{2}{3}\ln \left| \frac{t}{t_0} \right|\, ,\quad 
t_0^2 \equiv \frac{4}{3\kappa^2 \rho_0}\, .
\ee
Then for the finite $b$, by writing $N=\frac{2}{3}\ln \frac{t}{t_0} + \delta N$ and by 
using Eq.~(\ref{MKSN22}), we find 
\be
\label{MKSN24}
\delta\ddot N + \frac{4}{t} \delta \dot N + \frac{1}{t} \delta N
= \mathcal{O} \left( b^2 \right) + \mathcal{O} \left( \left(\delta N\right)^2 \right) \, ,,
\ee
whose solution is given by 
\be
\label{MKSN25}
\delta N = C_+ \left|t\right|^{\frac{-3 + \sqrt{7}}{2}} 
+ C_- \left|t\right|^{\frac{-3 - \sqrt{7}}{2}} 
+ \mathcal{O} \left( b^2 \right)\, .
\ee
Here $C_\pm$ are arbitrary constants. 
Because the first and the second terms do not depend on $b$, we may put $C_\pm=0$. 
In fact if we keep $C_+$, we find $E$ diverges and therefore physically not acceptable. 
On the other hand, even if we keep $C_-$, this term does not contribute to $E$. 
Then for the large $N$, by using the expression of $E$ in Eq.~(\ref{MKSN17}) with 
Eq.~(\ref{MKSN18}), we find
\be
\label{MKSN26}
E = - \frac{\left( b \kappa^2 \rho_0 \right)^2}{16b}\, .
\ee
Therefore when $b>0$, the shrinking of the universe will always stop and turn to 
expand, that is, we obtain the bouncing universe. 
On the other hand, when $b<0$, the shrinking universe always reaches 
the singular point at $1 + b \kappa^2 \rho_0 \e^{-3N}=0$. 

When $b>0$, we may estimate $N$ when the shrinking universe turns to expand 
and therefore $V(N)=E$. 
When $b\kappa^2 \rho_0 \gg 1$, by using the expression of $V(N)$ in Eq.~(\ref{MKSN19}) and 
$E$ in Eq.~(\ref{MKSN26}), we find the minimum $N_\mathrm{min}$ of $N$,  
\be
\label{MKSN27}
\e^{3N_\mathrm{min}} \sim \frac{3}{8} b\kappa^2 \rho_0\, .
\ee
On the other hand, when $b\kappa^2 \rho_0 \ll 1$, by using Eq.~(\ref{MKSN20}), we find 
\be
\label{MKSN28}
\e^{3N_\mathrm{min}} \sim \frac{9}{64} b\kappa^2 \rho_0\, .
\ee

We should note that Eq.~(\ref{MKSN17}) can be identified with the first FRW equation 
because $H=\dot N$ and rewritten as
\be
\label{FRW1}
\frac{3}{\kappa^2} H^2 = \frac{6}{\kappa^2} \e^{-6N} \left( 1 + \frac{b\kappa^2 \rho_0}{4}
\e^{-3N} \right)^2 \left( E - V(N) \right)\, .
\ee
For large $N$, the r.h.s. in Eq.~(\ref{FRW1}) can be expanded as a power series with respect to 
$\e^{-3N}$ and we find 
\be
\label{FRW2}
\frac{3}{\kappa^2} H^2 = \rho \left( 1 - \frac{\rho}{\rho_l} \right) + \mathcal{O}\left( 
\e^{-9N} \right)\, , \quad \rho = \rho_0 \e^{-3N}\, ,\quad 
\rho_l \equiv \frac{2}{b\kappa^2}\, .
\ee
The above structure is similar to that in the loop quantum cosmology 
\cite{Bojowald:1999tr,Bojowald:1999ts,Ashtekar:2006es}. 
In case of the loop quantum cosmology, instead of Eq.~(\ref{FRW2}), we have
\be
\label{loopFRW}
\frac{3}{\kappa^2} H^2 = \rho \left( 1 - \frac{\rho}{\rho_c} \right) \, .
\ee
Here $\rho_c$ is the critical density and the energy density $\rho$ is always equal to or smaller than 
$\rho_c$, $\rho\leq \rho_c$. 
In the loop quantum cosmology, the shrinking universe turns to expand when $\rho=\rho_c$. 
Even in our model, 
if we define the critical density $\rho_c^\mathrm{BI}$ by the density satisfying $E=V(N)$ in Eq.~(\ref{FRW1}), 
the shrinking universe turns to expand when $\rho=\rho_c^\mathrm{BI}$ but we find 
$\rho_c^\mathrm{BI} \neq \rho_l$ in Eq.~(\ref{FRW1}) due to the correction of $\mathcal{O}\left( \e^{-9N} \right)$ and 
by using Eq.~(\ref{FRW1}) or (\ref{FRW2}), we obtain $\rho_c^\mathrm{BI}$ as follows,  
\be
\label{FRW3}
\rho_c^\mathrm{BI} = \left\{ 
\begin{array}{ll}
\frac{8}{3b\kappa^2} & \mbox{when} \ b\kappa^2 \rho_0 \gg 1 \\
\frac{64}{9b\kappa^2} & \mbox{when} \ b\kappa^2 \rho_0 \ll 1 
\end{array} 
\right. \, .
\ee
Therefore the obtained behavior of the bouncing is similar to that in the loop quantum gravity, 
although there are quantitative differences.

\section{Black hole formation by the collapse of dust}\label{bhf}

In the last section, we have concluded that the FRW Universe filled with dust would bounce in the Born-Infeld gravity with positive $b$.
On the other hand, the result could be applied to a gravitational collapse of uniform and spherical ball of dust 
as in \cite{Oppenheimer:1939}. 
In this section, we consider if black hole can be formed by the collapse of dust. 
We now assume there is a spherically symmetric and uniform ball made 
of dust and consider the collapse of ball. 
In the Einstein gravity, this assumption is valid because the pressure of the dust vanishes. 
If the falling matter fluid has a pressure, the density of ball cannot be 
uniform because the pressure should vanish at the boundary between the 
ball and bulk, which is assumed to be vacuum. 
This assumption can be justified in the Einstein gravity by using the junction conditions. 
In the Born-Infeld gravity in the Palatini formalism, however, we do not know the exact expressions of 
the junction condition. 
Then the arguments below might not be justified quantitatively but we may expect that the results 
obtained in this section could be correct qualitatively.    

If we can regard that the space-time inside the ball of dust could be the shrinking FRW universe as in the last section, the results in the last section could tell that there would be a bouncing. 
If the radius of the ball at the bouncing is larger than the Schwarzschild 
radius, the black hole cannot be formed. 

We assume the ball of dust with radius $R$ at $N=N_0$. 
We choose $N_0$ to be large enough. Then the total mass $M$ is given by
\be
\label{MKSN29}
M = \frac{4\pi}{3} R^3 \rho_0 \e^{-3N_0}\, .
\ee
Here $\rho_0 \e^{-3N_0}$ is the energy density of the ball at $N=N_0$.  
We now consider the case that $b>0$. 
First we assume 
\be
\label{MKSN30}
b \kappa^2 \rho_0 = \frac{3 b \kappa^2 M \e^{3N_0}}{4\pi R^3} \gg 1\, .
\ee
Then by using Eq.~(\ref{MKSN27}), we find $N=N_b$ at the bouncing is given by
\be
\label{MKSN31}
\e^{3N_b} \sim \frac{9 b \kappa^2 M \e^{3N_0}}{32\pi R^3} \, ,
\ee
which give the radius $R_b$ at the bouncing by 
\be
\label{MKSN32}
R_b^3 = R^3 \e^{3 \left( N_b - N_0 \right)} 
= \frac{9 b \kappa^2 M}{32\pi}\, .
\ee
On the other hand, the Schwarzschild radius $R_s$ is given by 
\be
\label{MKSN33}
R_s = \frac{\kappa^2 M}{4\pi} \, .
\ee
Then we find
\be
\label{MKSN34}
\frac{R_b^3}{R_s^3} = \frac{18 \pi^2 b }{\kappa^4 M^2}\, .
\ee
Therefore large black hole, where $M^2\gg \frac{b}{\kappa^4}$, can be formed because $R_b \ll R_s$ 
and therefore the bouncing can occur after the formation of the horizon. 

Instead of Eq.~(\ref{MKSN30}), we may also consider the case 
\be
\label{MKSN35}
b \kappa^2 \rho_0 = \frac{3 b \kappa^2 M \e^{3N_0}}{4\pi R^3} \ll 1\, .
\ee
Then by using Eq.~(\ref{MKSN28}), we find that the bouncing occurs when 
\be
\label{MKSN36}
\e^{3N} \sim \e^{3\tilde N_b} \sim \frac{27 b \kappa^2 M \e^{3N_0}}{256\pi R^3} \, ,
\ee
and the radius $\tilde R_b$ at the bouncing is given by 
\be
\label{MKSN37}
R_b^3 = R^3 \e^{3 \left( N_b - N_0 \right)} 
= \frac{27 b \kappa^2 M}{256\pi}\, ,
\ee
and we obtain
\be
\label{MKSN38}
\frac{R_b^3}{R_s^3} = \frac{27 \pi^2 b }{4 \kappa^4 M^2}\, .
\ee
Therefore small black hole, where $M^2\ll \frac{b}{\kappa^4}$, cannot be formed because $R_b \gg R_s$. 

We now consider the case that $b<0$. 
In this case, there is a maximum $\rho_\mathrm{max}$ in the energy density $\rho$ given by Eq.~(\ref{MK14}). 
We now consider the meaning of the density $\rho_\mathrm{max}$ in the black hole formation. 
We now assume that the black hole is formed by the collapse of the star made of the dust 
with radius $r$. Then the energy density $\rho$ is given by 
\be
\label{MK18} 
\rho = \tilde\rho_0 r^{-3}\, .
\ee
Here $\tilde\rho_0$ is a constant. 
Then the mass $M$ and the Schwarzschild radius $R_s$ of the star is given by
\be
\label{MK19}
M= \frac{4\pi}{3} \rho r^3 = \frac{4\pi}{3} \tilde\rho_0\, ,\quad 
R_s = \frac{\kappa^2 M}{4\pi} = \frac{\kappa^2 \tilde\rho_0}{3}\, .
\ee
Then Eq.~(\ref{MK14}) tells that the minimum of $r$ is given by 
\be
\label{MK20}
r_\mathrm{min} = \left( - 3 b R_s \right)^{\frac{1}{3}}\, .
\ee
The black hole cannot be formed if $r_\mathrm{min}>R_s$, that is
\be
\label{MK21}
R_s^2 < - 3b\, .
\ee
Therefore small black holes may be prohibited if $b<0$ but  large ones are not prohibited. 
This result may tell that the creation of the primordial black holes might be prohibited. 

An important question is how the horizon formation could be consistent with the bouncing of the 
universe. 
In case of the loop quantum gravity, an expectation is the formation of the inner horizon as 
in the Reissner-Nordstrom space-time. 
The ball of the dust goes through the inner horizon and the ball might appear in another world 
as in the space-time of the Reissner-Nordstrom black hole or the Kerr black hole. 
Another possibility could be that the star which has bounced come back to the asymptotic region
through the regular black hole structure such as discussed in \cite{Hayward:2005gi}
At present, we do not have any definite answer and this question could be a future problem. 

In more realistic way, the above problem about the consistency between the horizon formation and 
the bouncing of the universe might be solved by the junction condition. 
In fact, there is an application of the Born-Infeld gravity to study the spherical symmetric compact star \cite{Cardoso:2012} by
using the junction conditions, where the Darmois-Israel formulation \cite{Israel:1966} has been used. 
We cannot, however, consider the junction conditions in a similar way because the second junction condition could not be 
formulated in our case.
The effective repulsive force which causes the bouncing universe makes the dust to move to the outside of the star
and then violates the homogeneity we assumed.
This inhomogeneity could be realized by introducing the shell on the boundary of the star.
Therefore there should appear the surface term proportional to a $\delta$-function in  
the energy-momentum tensor. 
Furthermore, because the surface term depends on time during the collapse, we cannot use simple shell model which has a constant tension. 
Therefore it could be a difficult but essential task to construct an appropriate model and solve the dynamics of the star.

We do not know the exact expression of the junction condition but 
we could be abe to use the junction condition in the Einstein gravity as an approximation. 
Therefore although the obtained results could not be completely justified, 
we may give the following speculation: 
The dust on the boundary between the sphere of the dust and the vacuum may move by the geodesic 
of the Schwarzschild space-time. 
Inside the sphere, there appears the effective repulsive force between the dust due to the correction 
by the Born-Infeld gravity, which may lead to the bouncing in the FRW universe. 
As mentioned above, the repulsive force makes the dust inhomogeneous and pushes the dust which was inside of the sphere,  
out of the boundary.  
The dust pushed out from the boundary will move by the geodesic of the Schwarzschild space-time but 
the geodesic is not always infalling one but going outward. 
Therefore if the bouncing occurs before the dust goes inside the horizon, there will occur the bouncing 
and the black hole might not be formed. 
On the other hand, even if the bouncing occurs inside the horizon, the dust may spread in the time 
direction, which is space like inside the horizon and the dust may not appear outside the horizon.

\section{Summary}

In the Born-Infeld gravity by using the Palatini formalism, we have investigated the FRW cosmology where the 
matter is dust and we have shown that when $b>0$, there occurs the bouncing. 
The cosmology in the Palatini-Born-Infeld gravity has been investigated in several 
papers, but in the most of the previous works, the connections are assumed to be 
given by $P_{\mu\nu}$, which resembles the metric of the FRW universe but 
this requirement is too strong and we considered more general case. 
By applying the results in the FRW universe, we also gave some speculations about the collapse of the 
sphere of dust and the black hole formation. 
Then we have shown that although the large black hole might be formed but the small 
black holes could be prohibited to be formed. 
This naive speculation about the singularity avoidance requires more quantitative 
arguments and we need to solve the junction conditions for the metric and connection so that 
the internal FRW space-time connects smoothly to the external Schwarzschild space-time. 
There is an application to the spherically symmetric and static compact star in the Palatini 
formalism by using the Darmois-Israel junction condition. 
In our case, however, there is another kind of difficulty. 
As mentioned in the last section, we have assumed that the dust could be homogeneous 
but because there appears the repulsive force 
between the dust in the Born-Infeld gravity, the dust becomes inhomogeneous during the collapse when 
there is a boundary between the dust and vacuum. 
In fact, if we solve the equations by assuming that the dust is homogeneous inside the boundary, there 
should appear a kind of shell, where the energy-momentum tensor diverges as a $\delta$-function. 
If we assume the junction condition in the Einstein gravity as an 
approximation, the approximation could be valid when the density of the dust is small and the total mass is 
large enough because the equations in the Born-Infeld gravity coincide with the Einstein equations in the 
limit that the energy-momentum tensor vanishes, that is, the energy density goes to zero. 
Furthermore, the repulsive force could be weak when the curvature at the horizon is small enough 
because the corrections from the Einstein gravity become small when $bR \ll 1$. 
Therefore the junction condition in the Einstein gravity can be a leading order approximation of the full 
junction condition in the Born-Infeld gravity in the Palatini formalism although the full and exact analysis 
could be complicated.
We will consider this problem in the future work. 

It could be interesting to investigate the possibility that a spherically symmetric space-time without 
singularity,  as is expected to appear in the final stage of dust collapse, may be an exact solution in the 
Born-Infeld gravity.
We should note that some regular black hole solutions are already known
\cite{Bardeen1968, Hayward:2005gi, Flachi:2012nv}. 
In the Einstein gravity, those solutions are not vacuum solutions, but a non-linear electromagnetic source 
is responsible to obtain the regular black holes 
\cite{AyonBeato:2000zs, Bronnikov:2000vy, Berej:2006cc}.
On the other hand, we found that the Born-Infeld gravity in the Palatini formalism may have the black 
hole solution without singularity from the speculation about the dust collapse.
If so, the non-linearity in gravitational action itself would work as the origin to remove the singularity
as in case of the non-linear electromagnetic source.
Additionally, if we could find the static configuration of the non-singular space-time with matter fields,
it could be an great interest to study the stability, the equation of state of matter inside the black hole, 
and the energy condition \cite{Dymnikova:2004zc}.

\section*{Acknowledgements}

The work by S.N. is supported by the JSPS Grant-in-Aid for Scientific
Research (S) \# 22224003 and (C) \# 23540296.
The work by T.K is supported by the Grant-in-Aid for JSPS Fellows \# 15J06973

\appendix

\section{Derivations of Eqs.~(\ref{MK9}), (\ref{MKSN5}), (\ref{MKSN7}), and (\ref{MKSN8}) 
\label{derivation}}

In this appendix we derive Eqs.~(\ref{MK9}), (\ref{MKSN5}), (\ref{MKSN7}), and (\ref{MKSN8}). 
By assuming Eq.~(\ref{MK1}) and (\ref{MK2}), we find that the Ricci tensors are given by 
\be
\label{MK3} 
R_{tt} = - 3 \left(\dot C + C^2 - AC \right)\, , \quad 
R_{ij} = a^2 \left( \dot B + 2HB + BC + BA \right) \delta_{ij}\, , \quad 
R_{ti} = R_{it} = 0\, ,
\ee
Then we obtain the following equations:
\begin{align}
\label{MK4}
b \kappa^2 \rho = & \left\{ 1 + b \left( \dot B + 2HB + BC + BA \right) \right\}^{\frac{3}{2}} 
\left\{ 1 + 3b \left(\dot C + C^2 - AC \right) \right\}^{- \frac{1}{2}} - 1 \, ,\\
\label{MK5}
0 = & \left\{ 1 + b \left( \dot B + 2 HB + BC + BA \right) \right\}^{\frac{1}{2}} 
\left\{ 1 + 3b \left(\dot C + C^2 - AC \right) \right\}^{1 \frac{1}{2}} - 1 \, ,\\
\label{MK6}
\Gamma^t_{tt} = & \frac{1}{2} \frac{d}{dt} \left\{ \ln \left\{ 1 + 3b \left(\dot C 
+ C^2 - AC \right) \right\} \right\} \, , \\
\label{MK7} 
\Gamma^t_{ij} =& \frac{a(t)^2}{2} 
\left\{ 1 + 3 b \left(\dot C + C^2 - AC \right) \right\}^{-1} \nn
& \times \left\{ 2 H + b \left\{ 4 H \dot B + 4 H^2 B + 2 HBC + 2 HBA + \ddot B 
+ 2 \dot H B 
+ \dot B \left( C + A  \right) + B \left( \dot C + \dot A \right) \right\} \right\}\delta_{ij} 
\, , \\
\label{MK8}
\Gamma^i_{jt} = \Gamma^i_{tj} = & \frac{1}{2} \frac{d}{dt} \left\{ \ln \left\{ a^2 
+ b a^2 \left( \dot B + 2HB + BC + BA \right) \right\} \right\}\delta^i_{\ j} \, .
\end{align}
Eq.~(\ref{MK4}) is the $(t,t)$ component of (\ref{eq:metric}) and Eq.~(\ref{MK5}) is the $(i,j)$ component. 
Eqs.~(\ref{MK6}), (\ref{MK7}), and (\ref{MK8}) are solutions of Eq.~(\ref{eq:connection}). 

By using Eqs.~(\ref{MK2}), (\ref{MK5}), (\ref{MK6}), and (\ref{MK8}), we find the first equality $A =C-H$ in Eq.~(\ref{MK9}). 
We may delete $A$ by using Eq.~(\ref{MK9}) and obtain
\begin{align}
\label{MK10}
b \kappa^2 \rho = & \left\{ 1 + b \left( \dot B + 3HB \right) \right\}^{\frac{3}{2}} 
\left\{ 1 + 3b \left(\dot C + 2 C^2 - CH \right) \right\}^{- \frac{1}{2}} - 1 \, ,\\
\label{MK11}
0 = & \left\{ 1 + b \left( \dot B + 3HB \right) \right\}^{\frac{1}{2}} 
\left\{ 1 + 3b \left(\dot C + 2 C^2 - CH \right) \right\}^{\frac{1}{2}} -1\, ,\\
\label{MK12}
 - C + H = & \frac{1}{2} \frac{d}{dt} \left\{ \ln \left\{ 1 + 3b \left(\dot C + 2 C^2 - CH 
\right) \right\} \right\} \, , \\
\label{MK13} 
B =& \left\{ 1 + 3 b \left(\dot C + 2 C^2 - CH \right) \right\}^{-1} \nn
& \times \left\{ 2 H + b \left\{ 5 H \dot B + 6 H^2 B + 3 \dot H B + \ddot B \right\} 
\right\} \, .
\end{align}
Furthermore by using Eqs.~(\ref{MK10}) and (\ref{MK11}), we obtain
\be
\label{MK15}
b \kappa^2 \rho = \left\{ 1 + b \left( \dot B + 3HB \right) \right\}^2 - 1 \, .
\ee
We now delete $B$ and $C$ in Eqs.~(\ref{MK10}), (\ref{MK11}), (\ref{MK12}), (\ref{MK13}) and 
obtain a single equation with respect to the scale factor $a$. 
By assuming Eq.~(\ref{MKSN1}) and by combining Eqs.~(\ref{MK11}) and (\ref{MK13}), we obtain
\be
\label{MKSN3}
4 a^4 B = \frac{d}{dt} \left[ a^4 \left\{ 1 + b \left( \dot B + 3HB \right) 
\right\}^2 \right]\, .
\ee
Furthermore by combining (\ref{MK15}) and (\ref{MKSN3}), we find (\ref{MKSN5}). 
On the other hand, Eqs.~(\ref{MK10}) and (\ref{MK11}) give 
\be
\label{MKSN6}
b \kappa^2 \rho = \left\{ 1 + 3b \left(\dot C + 2 C^2 - CH \right) \right\}^{- 2} - 1\, .
\ee
By using Eqs.~(\ref{MK12}) and (\ref{MKSN6}), we obtain Eq.~(\ref{MKSN7}) and the second equality (\ref{MK9}). 
A single equation (\ref{MKSN8}) with respect to the scale factor $a$ can be obtained 
by deleting $B$ in Eq.~(\ref{MK15}) by using Eq.~(\ref{MKSN5}).


\begin{thebibliography}{99}

\bibitem{review}
S.~Nojiri and S.~D.~Odintsov,
Phys.\ Rept.\  {\bf 505} (2011) 59
[arXiv:1011.0544 [gr-qc]]; \\
S.~Nojiri and S.~D.~Odintsov,
eConf C {\bf 0602061} (2006) 06
[Int.\ J.\ Geom.\ Meth.\ Mod.\ Phys.\  {\bf 4} (2007) 115]
[hep-th/0601213]; \\
S.~Nojiri and S.~D.~Odintsov,
Int.\ J.\ Geom.\ Meth.\ Mod.\ Phys.\  {\bf 11} (2014) 1460006
[arXiv:1306.4426 [gr-qc]]; \\
K.~Bamba, S.~Capozziello, S.~Nojiri and S.~D.~Odintsov,
Astrophys.\ Space Sci.\ {\bf 342} (2012) 155
[arXiv:1205.3421 [gr-qc]]; \\
S.~Capozziello and M.~De Laurentis,
Phys.\ Rept.\ {\bf 509} (2011) 167
[arXiv:1108.6266 [gr-qc]]; \\
V.~Faraoni and S.~Capozziello,
Fundamental Theories of Physics, Vol. 170, Springer, 2010; \\
A.~Joyce, B.~Jain, J.~Khoury and M.~Trodden,
arXiv:1407.0059 [astro-ph.CO]; \\
A.~de la Cruz-Dombriz and D.~Saez-Gomez,
Entropy {\bf 14} (2012) 1717
[arXiv:1207.2663 [gr-qc]].


\bibitem{Deser:1998rj}
S.~Deser and G.~W.~Gibbons,
Class.\ Quant.\ Grav.\  {\bf 15} (1998) L35
[hep-th/9803049].

\bibitem{Vollick:2003qp}
D.~N.~Vollick,
Phys.\ Rev.\ D {\bf 69} (2004) 064030
[gr-qc/0309101].

\bibitem{Born:1934gh}
M.~Born and L.~Infeld,
Proc.\ Roy.\ Soc.\ Lond.\ A {\bf 144} (1934) 425.



\bibitem{Eddington}
A.~S.~Eddington, ``The mathematical theory of relativity'', 
(Cambridge University Press, Cambridge, England. 1924).









\bibitem{Banados:2010ix}
M.~Banados,
Phys.\ Rev.\ D {\bf 77} (2008) 123534
[arXiv:0801.4103 [hep-th]]; \\
%
M.~Banados and P.~G.~Ferreira,
Phys.\ Rev.\ Lett.\  {\bf 105} (2010) 011101
[arXiv:1006.1769 [astro-ph.CO]]; \\
J.~H.~C.~Scargill, M.~Banados and P.~G.~Ferreira,
Phys.\ Rev.\ D {\bf 86} (2012) 103533
[arXiv:1210.1521 [astro-ph.CO]].


\bibitem{Avelino:2012ue}
P.~P.~Avelino and R.~Z.~Ferreira,
Phys.\ Rev.\ D {\bf 86} (2012) 041501
[arXiv:1205.6676 [astro-ph.CO]].






\bibitem{Chemissany:2008fy}
W.~A.~Chemissany, M.~de Roo and S.~Panda,
Class.\ Quant.\ Grav.\  {\bf 25} (2008) 225009
[arXiv:0806.3348 [hep-th]].



\bibitem{Bojowald:1999tr}
M.~Bojowald,
Class.\ Quant.\ Grav.\  {\bf 17} (2000) 1489
[gr-qc/9910103].

\bibitem{Bojowald:1999ts}
M.~Bojowald,
Class.\ Quant.\ Grav.\  {\bf 17} (2000) 1509
[gr-qc/9910104].

\bibitem{Ashtekar:2006es}
A.~Ashtekar, T.~Pawlowski, P.~Singh and K.~Vandersloot,
Phys.\ Rev.\ D {\bf 75} (2007) 024035
[gr-qc/0612104].

\bibitem{Nojiri:1999ji}
S.~Nojiri and S.~D.~Odintsov,
Phys.\ Lett.\ B {\bf 471} (1999) 155
[hep-th/9908065].



\bibitem{Cardoso:2012}
P.~Pani, T.~Delsate, and V.~Cardoso, 
Phys.\ Rev.\ D {\bf 85}, 084020 (2012)

\bibitem{Israel:1966}
W.~Israel, 
Nuovo Cim {\bf B44S10}, 1 (1966)

\bibitem{Oppenheimer:1939}
J.~R.~Oppenheimer and H.~Snyder, 
Phys.\ Rev.\ {\bf 56, 455} (1939) 

\bibitem{Bardeen1968}
J.~M.~Bardeen, 
Proceedings of GR5 (Tbilisi, USSR, 1968), p. 174.

\bibitem{Hayward:2005gi}
S.~A.~Hayward,
Phys.\ Rev.\ Lett.\  {\bf 96} (2006) 031103
[gr-qc/0506126].

\bibitem{Flachi:2012nv}
A.~Flachi and J.~P.~S.~Lemos,
Phys.\ Rev.\ D {\bf 87} (2013) 2,  024034
[arXiv:1211.6212 [gr-qc]].

\bibitem{AyonBeato:2000zs}
E.~Ayon-Beato and A.~Garcia,
Phys.\ Lett.\ B {\bf 493} (2000) 149
[gr-qc/0009077].

\bibitem{Bronnikov:2000vy}
K.~A.~Bronnikov,
Phys.\ Rev.\ D {\bf 63} (2001) 044005
[gr-qc/0006014].

\bibitem{Berej:2006cc}
W.~Berej, J.~Matyjasek, D.~Tryniecki and M.~Woronowicz,
Gen.\ Rel.\ Grav.\  {\bf 38} (2006) 885
[hep-th/0606185].

\bibitem{Dymnikova:2004zc}
I.~Dymnikova,
Class.\ Quant.\ Grav.\  {\bf 21} (2004) 4417
[gr-qc/0407072].




\end{thebibliography}
\end{document}